\documentclass[conference]{IEEEtran}
\usepackage{graphicx,siunitx}
\graphicspath{{./figures/},{./imgs/}}
\usepackage{booktabs,multirow,longtable}
\usepackage{algorithm,algorithmic}
\usepackage{balance}
\usepackage{enumitem}
\usepackage[cmex10]{amsmath}
\usepackage[utf8x]{inputenc}
\usepackage[T1]{fontenc}
\usepackage[american]{babel}
\usepackage{stfloats}
\usepackage{amsthm, amssymb, bm, bbm, graphicx, ucs, cite, relsize, booktabs,color,, mathalfa, upgreek, tablefootnote, multirow, colortbl}
\usepackage[hidelinks]{hyperref}

\usepackage{fancyhdr}
\IEEEoverridecommandlockouts
\fancypagestyle{myfancy}{
	\fancyhf{} 
	\fancyhead[C]{\textit{Accepted to the 33rd European Signal Processing Conference (EUSIPCO 2025), Isola delle Femmine, Palermo, Italy}} 
	\fancyfoot[C]{\footnotesize © 2025 IEEE. Personal use of this material is permitted. 
		Permission from IEEE must be obtained for all other uses.}

}

\definecolor{Gray}{gray}{0.9}

\theoremstyle{remark}

\theoremstyle{definition}

\theoremstyle{remark}

\newcommand*{\herm}{^{\mathsf{H}}}
\newcommand*{\transp}{^{\mathsf{T}}}

\DeclareMathOperator{\diag}{diag}

\DeclareMathOperator*{\argmax}{\arg\max}

\newcommand{\e}{\mathrm{e}}
\renewcommand{\i}{\mathrm{i}}

\IEEEoverridecommandlockouts
\allowdisplaybreaks

\newcommand{\test}{\mathrel{\underset{{\mathcal H}_{0}}{\overset{{\mathcal{H}}_{1}}{\gtrless}}}} 

\begin{document}

\title{Track-before-detect in RIS-aided Integrated Sensing and Communication}

\author{
\IEEEauthorblockN{Georgios Mylonopoulos$^{1,3,4}$, Luca Venturino$^{1,3,4}$, Emanuele Grossi$^{1,3,4}$, Stefano Buzzi$^{1,2,3,4}$, and Ciro D'Elia$^{1,3,4}$} 
\IEEEauthorblockA{$^1$\textit{University of Cassino and Southern Lazio, 03043 Cassino, Italy}\\
$^2$\textit{Polytechnic University of Milan, 02 Milan, Italy}\\
$^3$\textit{European University of Technology EUt+, European Union} \\
$^4$\textit{Consorzio Nazionale Interuniversitario per le Telecomunicazioni (CNIT), 43124 Parma, Italy}\\
\{georgios.mylonopoulos, l.venturino, e.grossi, s.buzzi, delia\}@unicas.it
\thanks{The work of G. Mylonopoulos and S. Buzzi was supported by the EU Horizon 2020 MSCA-ITN-METAWIRELESS, Grant Agreement 956256. The work of L. Venturino was supported by the project ``FLexible And distributed cognitive Radar systEms'' (FLARE), CUP E63C22002040007, funded by the European Union under the Italian National Recovery and Resilience Plan of NextGenerationEU, partnership on ``Telecommunications of the Future'' (Project PE00000001, program ``RESTART''). The work of E. Grossi was supported by the project ``CommunIcations and Radar Co-Existence (CIRCE),'' CUP H53D23000420006, funded by the European Union under the Italian National Recovery and Resilience Plan of NextGenerationEU.}
}
}

\maketitle

\thispagestyle{myfancy}

\begin{abstract}
This study considers a base station equipped with sensing and communication capabilities, which serves a ground user and scans a portion of the sky via a passive reconfigurable intelligent surface. To achieve more favorable system tradeoffs, we utilize a multi-frame radar detector, comprising a detector, a plot-extractor, and a track-before-detect processor. The main idea proposed here is that user spectral efficiency can be enhanced by increasing the number of scans jointly processed by the multi-frame radar detector while maintaining the same sensing performance. A numerical analysis is conducted to verify the effectiveness of the proposed solution and to evaluate the achievable system tradeoffs.
\end{abstract}


\section{Introduction}
Future wireless networks may leverage the same hardware and physical resources for both sensing and communication to improve power and spectral efficiency (SE), reduce carbon footprint and electromagnetic pollution, and implement novel applications relying on both functions~\cite{9606831,9737357,10012421,9585321}. Existing studies on integrated sensing and communication (ISAC) have already identified relevant performance metrics and developed various resource allocation strategies to capture the inherent tradeoffs between these two functions~\cite{9376324,9540344,9705498}. 
For example, the space-time waveform generated by the multi-antenna transmitter, the power allocation across time, frequency, and space, and the communication receiver's design can be optimized to maximize communication metrics such as SE, power efficiency, and signal-to-interference-plus-noise ratio (SINR). This optimization is performed under constraints related to sensing metrics, including the signal-to-clutter ratio, detection probability, and the accuracy of estimating unknown parameters, or vice versa.

In future network evolutions, a major challenge is the identification of cost-effective ISAC strategies that require limited modifications of existing infrastructures and protocols, and plug-and-play solutions that can be seamlessly integrated with existing technologies would be extremely attractive. In this context, a reconfigurable intelligent surface (RIS) may be easily incorporated into the design of ISAC systems; in particular, such technology could be used to better manage the interference between the sensing and communication functions, expand the field of view of existing dual-function base stations, or deploy RIS-based massive MIMO transceivers at low cost~\cite{10077119,Buzzi-2022,Tulino-2021,Grossi-2023,Grossi-2024}. On the other hand, advanced signal processing techniques developed by radar engineers might be included in the design of ISAC systems to obtain more favorable sensing and communication tradeoffs. For example, track-before-detect (TBD) procedures have been proven effective for detecting weak moving targets and estimating their position~\cite{Boers_2004,Blanding_2007,Davey_2008,Buzzi-2008,Pulford_2010}, as they are able to accumulate the energy back-scattered by a target with an unknown trajectory over multiple scans (or frames). In this context, a computationally efficient two-stage TBD processor has been developed in~\cite{Grossi-2013a, Grossi-2013b, Aprile-2016}, wherein the first stage limits the number of observations by retaining only the most significant ones, and the second stage performs trajectory formation and detection validation. 

In this study, we integrate the TBD-based multi-frame detector from~\cite{Grossi-2013a, Grossi-2013b, Aprile-2016} into an ISAC transceiver that accomplishes both downlink communication towards multiple ground users and RIS-aided sensing of airborne targets. More specifically, we propose introducing additional degrees of freedom for system design by exploiting the temporal correlation of the echoes generated by a moving target. Our results show that, by increasing the number of scans processed by the radar detector (and therefore its implementation complexity), we can reduce the amount of power dedicated to the sensing function while maintaining the same performance (measured in terms of probability of target detection and root mean square error in the estimation of the target position); this excess power can be reused to increase the user SE. A numerical example based on 5G system specifications is provided for illustration.

\begin{figure}[t]
    \centering
    \includegraphics[width=\columnwidth]{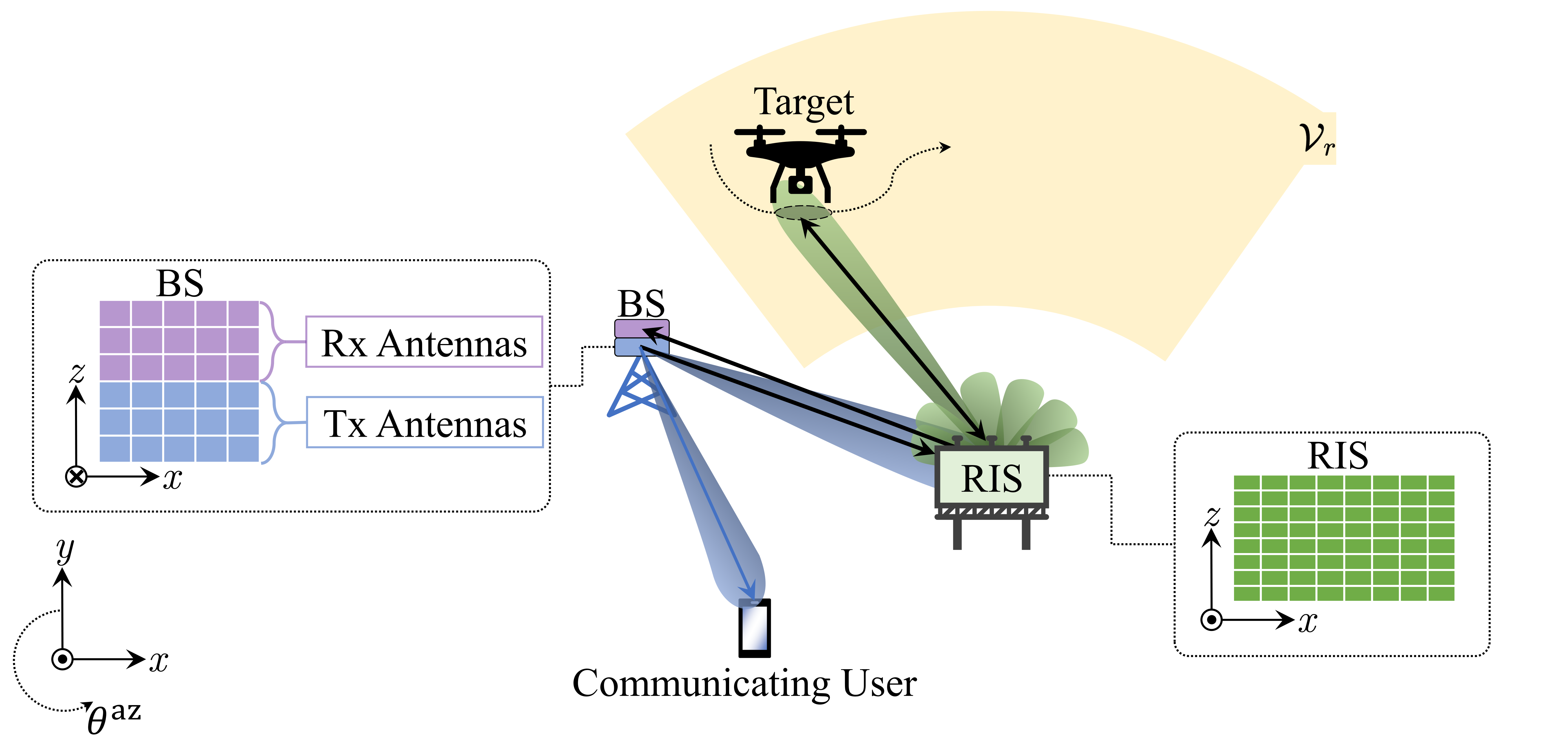}
    \caption{Considered system architecture. The BS simultaneously serves a single downlink ground user and, with the help of a passive RIS, senses an airborne target in the volume $\mathcal{V}_{s}$.}
    \label{fig:sm}
\end{figure}

\emph{Notation}: Column vectors (matrices) are denoted by lowercase (uppercase) boldface letters. The symbols $(\cdot)\transp$, $(\cdot)^*$ and $(\cdot)\herm$ denote the transpose, conjugate, and conjugate-transpose operations, respectively. $\bm{I}_M$ is the $M \times M$ identity matrix, while $\bm{0}_{M}$ is an $M \times 1$ vector of all zero elements. Finally, the symbols $\i$, $\varnothing$ and $\mathbb{E}\{\cdot\}$ denote the imaginary unit, an empty set and the statistical expectation, respectively.

\section{System description}\label{sec:system}
Consider a full-duplex BS equipped with a transmit planar array of $D_{\rm tx}$ antennas and one receive planar array of $D_{\rm rx}$ antennas. The BS adopts an OFDM transmission format with a central carrier frequency $f_{o}$, a number of subcarriers $N_{o}$, a subcarrier spacing $W_{o}$, and a cyclic prefix of duration $T_{o}$, whereby the duration of each OFDM symbol is $T_{\rm sym}=1/W_{o}+T_{o}$. A subset of $N_{\rm sub}$ subcarriers, equally-spaced by $W_{{\rm sub}}$, is employed to serve a single-antenna ground user and to detect a prospective airborne target in the volume $\mathcal{V}_{s}$, as shown in Fig.~\ref{fig:sm}. The corresponding subcarrier frequencies are $f_{1},\ldots,f_{N_{\rm sub}}$. The sensing function is implemented via a passive RIS with $D_{\rm ris}$ reflective elements arranged into a rectangular planar array. The RIS scans a volume not directly observable by the BS~\cite{Buzzi-2022}. We denote by $\mathcal{H}_{0}$ and $\mathcal{H}_{1}$ the hypotheses that no target and one target is present in the inspected volume $\mathcal{V}_{s}$, respectively.

Denote by $[R_{\min},R_{\max}]$, $[\theta_{\min}^{\rm{az}},\theta_{\max}^{\rm{az}}]$, and $[\theta_{\min}^{\rm{el}},\theta_{\max}^{\rm{el}}]$ the range, azimuth, and elevation intervals of the inspected volume $\mathcal{V}_{s}$ relative to the RIS, respectively. Also, let $\bm{\theta}= [\theta^{\rm az},\;\theta^{\rm el}]\transp$ be the angular direction specified by azimuth and elevation angles $\theta^{\rm az}$ and $\theta^{\rm el}$, respectively; then, we denote by $\bm{t}_{q}(\bm{\theta})$ the far-field steering vector of the RIS towards $\bm{\theta}$ on the $q$-th subcarrier, for $q=1,\ldots,N_{\rm sub}$. Upon partitioning $\mathcal{V}_{s}$ into $N_{\rm dir}$ subvolumes, we denote by $\bar{\bm{\theta}}_{i}=[\bar\theta_{i}^{\rm az}\;\bar\theta_{i}^{\rm el}]\transp$ the nominal pointing direction corresponding to the $i$-th subvolume and by $\mathcal{G}_{\rm dir}=\{\bar{\bm{\theta}}_{1},\ldots,\bar{\bm{\theta}}_{N_{\rm dir}}\}$ the set of all such directions. The RIS-assisted radar sequentially illuminates each subvolume for a coherent processing interval (CPI) containing $N_{\rm sym}$ OFDM symbols, say $T_{\rm cpi}=N_{\rm sym}T_{\rm sym}$, and it remains inactive for $B_{\rm sym}$ OFDM symbols between consecutive illuminations. 

\subsection{Transmit signal within a CPI} 
The signal emitted on the $q$-th subcarrier of the $n$-th symbol interval is
 \begin{equation}
\bm{p}_{q}(n)=\sqrt{\mathcal{P}}\Big(\bm{f}_{q,c}x_{q,c}(n)+\bm{f}_{q,s}x_{q,s}(n)\Big),
 \end{equation} 
for $q=1,\ldots,N_{\rm sub}$ and $n=1,\ldots,N_{\rm sym}$, where $\mathcal{P}$ is the power radiated on each subcarrier, while
$\bm{f}_{q,c}$ and $\bm{f}_{q,s}$ are the unit-norm beamformers corresponding to the communication and sensing symbols $x_{q,c}(n)$ and $x_{q,s}(n)$, respectively. The beamformers can be designed according to any of the criteria available in the literature (e.g., see~\cite{zf_methods}) and are assumed given in the following. The emitted symbols are assumed zero-mean and independent, with $\mathbb{E}[|x_{q,c}(n)|^2]=1-\gamma_{q}$ and $\mathbb{E}[|x_{q,s}(n)|^2]=\gamma_{q}$, where
$\gamma_{q}\in[0,1]$ is the fraction of power assigned to the sensing function. 

\subsection{Received signal within a CPI} 
Let $\bm{\omega}\in\Omega^{D_{\rm ris}}$ be the vector modeling the RIS response, with $\Omega=\{\chi\in\mathbb C: |\chi|=1\}$. Then,
the signal at the radar receiver on the $q$-th subcarrier in the $n$-th symbol interval is
\begin{subequations}\label{eq:rx_signal}
\begin{align}
\bm{y}_{q,s}(n)&=\bm{z}_{q,s}(n), &\text{under }\mathcal{H}_{0}, \label{eq:rx_signal_H0}\\
\bm{y}_{q,s}(n)&=\alpha\e^{-\i 2\pi q W_{\rm sub} \tau} \e^{-\i 2\pi \nu nT_{\rm sym}}\notag\\
&\quad \times \bm{G}_{q,\rm rx}\diag\{\bm{\omega}\}\bm{t}_{q}(\bm{\phi})\notag \\
&\quad \times \bm{t}_{q}\transp(\bm{\phi})\diag\{\bm{\omega}\}\bm{G}_{q,\rm tx}\bm{p}_{q}(n)\notag\\
&\quad +\bm{z}_{q,s}(n), &\text{under }\mathcal{H}_{1}, \label{eq:rx_signal_H1}
\end{align}
\end{subequations}
for $q=1,\ldots,N_{\rm sub}$ and $n=1,\ldots,N_{\rm sym}$, where: $\bm{z}_{q,s}(n)\in\mathbb{C}^{D_{\rm rx}}$ is a circularly-symmetric Gaussian vector with covariance matrix $\sigma^{2}_{s}\bm{I}_{D_{\rm rx}}$ accounting for the additive noise; $\bm{G}_{q,\rm tx}\in\mathbb{C}^{D_{\rm ris}\times D_{\rm tx}}$ and $\bm{G}_{q,\rm rx}\in\mathbb{C}^{D_{\rm rx}\times D_{\rm ris}}$ are the channel matrices between the BS transmitter and the RIS and between the RIS and the BS receiver, respectively; $\alpha$, $\tau$, $\nu$, $\phi^{\rm az}$, and $\phi^{\rm el}$ are the complex amplitude, delay, Doppler shift, azimuth angle, and elevation angle of the target, respectively; and, finally, $\bm{\phi}=[\phi_{\min}^{\rm{az}},\phi_{\max}^{\rm{az}}]$.

Upon assuming that $\bm{G}_{q,\rm rx}$ and $\bm{G}_{q,\rm tx}$ are constant over the CPI and known to the BS, we define the two-way (power) beampattern of the RIS-aided radar on the $q$-th subcarrier as
\begin{align}
	\text{BP}_{q}(\bm{\theta})	&= \frac{1}{\mathcal{P}}\mathbb{E}\Big[\big\|\bm{G}_{q,\rm rx}\diag\{\bm{\omega}\}\bm{t}_{q}(\bm{\theta})\notag \\
	&\quad \times \bm{t}_{q}\transp(\bm{\theta})\diag\{\bm{\omega}\}\bm{G}_{q,\rm tx}\bm{p}_{q}(n)\big\|^2\Big] \notag \\
	&= \Big[\|\bm{\omega}\transp \diag\{\bm{t}_{q}(\bm{\theta})\}\bm{G}_{q,\rm tx}\bm{f}_{q,c}\|^2(1-\gamma_{q})\notag \\
	&\quad +   \|\bm{\omega}\transp \diag\{\bm{t}_{q}(\bm{\theta})\}\bm{G}_{q,\rm tx}\bm{f}_{q,s}\|^2\gamma_{q}\Big] \notag \\
	&\quad \times  \big\|\bm{G}_{q,\rm rx}\diag\{\bm{t}_{q}(\bm{\theta})\}\bm{\omega}\big\|^2.
    \label{BP_per_sub}
\end{align}
We also define the two-way beampattern of the RIS-aided radar over the entire frequency band as
\begin{align}\label{eq:BP}
\text{BP}(\bm{\theta}) &= \sum_{q=1}^{N_{\rm sub}}\text{BP}_{q}(\bm{\theta}),
\end{align}
representing the normalized spatial power response from a prospective target located at angle $\bm{\theta}$ relative to the RIS.

Accordingly, we define the target signal-to-noise ratio (SNR) under $\mathcal{H}_{1}$ as
\begin{equation}
\mathrm{SNR}_{s}=\frac{N_{\rm{sym}}\mathcal{P}\,\text{BP}(\bm{\phi})\,\mathbb{E}\{|\alpha|^2\}}{\sigma^{2}_{s}}.
\label{eq:SNR_{s}}
\end{equation}
Note that the communication signal is not treated as interference here, as any transmitted signal known to the BS may be utilized to support the sensing function.

The signal received by the user on the $q$-th subcarrier in the $n$-th symbol interval is
\begin{align}
y_{q,c}(n) = \bm{h}_{q,c}\herm 
\bm{p}_{q}(n)+ z_{q,c}(n),\label{user-rx-signal}
\end{align}
for $q=1,\ldots,N_{\rm sub}$ and $n=1,\ldots,N_{\rm sym}$, where $\bm{h}_{q,c}\in\mathbb{C}^{D_{\rm tx}}$ is the channel between the BS and user on the $q$-th subcarrier, assumed constant over the CPI and known to the BS, and $z_{q,k}(n)\in\mathbb{C}$ is the additive noise, modeled as a circularly-symmetric Gaussian random variable with variance $\sigma^{2}_{c}$. Accordingly, the user SINR on the $q$-th subcarrier is
\begin{equation}
\mathrm{SINR}_{q,c}=\frac{(1-\gamma_{q}) |\bm{h}_{q,c}\herm \bm{f}_{q,c}|^2}{\displaystyle \gamma_{q} |\bm{h}_{q,c}\herm\bm{f}_{q,s}|^2+\frac{\sigma^{2}_{c}}{\mathcal{P}}},
\end{equation}
for $q=1,\ldots,N_{\rm sub}$, where the sensing signal is now treated as interference at the user. Also, assuming Gaussian signaling, the achievable SE (measured in bit/s/Hz) is
\begin{equation}\label{eq:se}
\mathrm{SE}=\frac{1}{T_{\rm sym} N_{\rm{sub}} W_{o}}\sum_{q=1}^{N_{\rm{sub}}}\log_{2}(1+\mathrm{SINR}_{q,c}).
\end{equation}

\section{Proposed system design}\label{sec:system_design}
Both the two-way beampattern of the RIS-aided radar in~\eqref{eq:BP} and the user SE in~\eqref{eq:se} depend on the fraction of power allocated to the sensing and communication functions; by increasing $\gamma_{q}$, the amount of power redirected towards the target is increased at the price of reducing the user SINR on the $q$-th subcarrier. Different sensing and communication tradeoffs can be obtained by varying $\{\gamma_{q}\}_{q=1}^{N_{\rm{sub}}}$, as discussed in Sec.~\ref{sec:Numerical analysis}.

On the other hand, the RIS response $\bm{\omega}$ and the implementation of the radar detector do not impact the communication function; therefore, they can be specifically designed to improve the sensing performance. Assuming absence of clutter, the RIS response can be chosen to maximize the amount of power redirected towards the subvolume under inspection. For the $i$-th pointing direction, the considered design problem is
\begin{align}\label{eq:BP_problem_formulation_max}
\max_{\bm{\omega}\in\Omega^{D_{\rm ris}}}& \text{BP}\left(\bar{\bm{\theta}}_{i};\bm{\omega}, \{\tilde{\bm f}_{q,s}\}_{q=1}^{N_{\rm{sub}}}\right), 
\end{align}
which is non-convex. A suboptimal solution to \eqref{eq:BP_problem_formulation_max} is found via an alternate maximization over the entries of $\bm{\omega}$. As to the radar detector, we propose to elaborate multiple, say $N_{\rm scan}$, consecutive scans to extend the time-on-target, as discussed in Sec.~\ref{sec:TBD}. This latter point appears particularly attractive since a larger complexity of the radar detector (i.e., a larger value of $N_{\rm scan}$) can be traded for a smaller value of $\gamma_{q}$ while maintaining the same sensing performance and improving the communication performance, as demonstrated in Sec.~\ref{sec:Numerical analysis}.

\subsection{Design of the radar detector}\label{sec:TBD}
The measurements collected in each CPI are processed through a bank of correlators, each matched to a delay, say $\tau$, and a Doppler shift, say $\nu$, taken from the uniformly-spaced grids $\mathcal{G}_{\rm del}=\{\bar{\tau}_1,\ldots,\bar{\tau}_{N_{\rm del}}\}$ and $\mathcal{G}_{\rm dop}=\{\bar{\nu}_1,\ldots,\bar{\nu}_{N_{\rm dop}}\}$, respectively; each correlator is followed by a square law detector and a noise variance normalization. For any $(\bm{\theta},\tau,\nu)\in \mathcal{G}_{\rm dir}\times \mathcal{G}_{\rm del}\times \mathcal{G}_{\rm dop}$, the output statistic is
\begin{multline}\label{eq:filter_output}
A(\bm{\theta},\tau,\nu) =\frac{1}{\sigma_{s}^{2}}\left|\sum_{q=1}^{N_{\rm sub}}\sum_{n=1}^{N_{\rm sym}} \bm{u}_{q}\herm(n;\bm{\theta},\tau,\nu) \bm{y}_{q,s}(n)\right|^2,
\end{multline}
where, from~\eqref{eq:rx_signal_H1}, the correlator $\bm{u}_{q}(n;\bm{\theta},\tau,\nu)$ is designed as
\begin{align}
\bm{u}_{q}(n;\bm{\theta},\tau,\nu)&=\e^{-\i 2\pi q W_{\rm sub} \tau}\e^{-\i 2\pi \nu nT_{\rm sym}}\notag \\
&\quad \times\frac{\bm{G}_{q,\rm rx}\diag\{\bm{\omega}\}\bm{t}_{q}(\bm{\theta})}{\|\bm{G}_{q,\rm rx}\diag\{\bm{\omega}\}\bm{t}_{q}(\bm{\theta})\|}\notag \\
&\quad \times \frac{\bm{t}_{q}\transp(\bm{\theta})\diag\{\bm{\omega}\}\bm{G}_{q,\rm tx}\bm{p}_{q}(n)}{|\bm{t}_{q}\transp(\bm{\theta})\diag\{\bm{\omega}\}\bm{G}_{q,\rm tx}\bm{p}_{q}(n)|}.\label{eq:filter}
\end{align}
In the $\ell$-th scan, the statistics in~\eqref{eq:filter_output} are collected into the array $\bm{A}_{\ell}$ of size $N_{\rm dir} \times N_{\rm del}\times N_{\rm dop}$, wherein the entry $(i,j,d)$ corresponds to the pointing direction $\bar{\bm{\theta}}_{i}$, the delay $\bar{\tau}_{j}$, and the Doppler shift $\bar{\nu}_{d}$. $\bm{A}_{\ell}$ is then sent to the \emph{detector and plot-extractor}, which produces a list of measurements called \emph{plots}. This block can be modeled as a multi-peak detector, where a plot is formed for each peak in $\bm{A}_{\ell}$ exceeding a given threshold $\eta_{\rm plot}$. Here, a plot is defined as a 5-dimensional row vector containing the value of the test statistic in~\eqref{eq:filter_output}, the position (range, azimuth, and elevation) of the peak, and the time at which the subvolume containing the detected peak has been illuminated. The plots are organized in the plot-list $\bm S_\ell$, which is a 5-column matrix with the extracted plots as rows, and sent to the TBD processor, as in~\cite{Grossi-2013a}; if no plot is formed, $\bm S_\ell =\emptyset$.

The TBD processor exploits the correlation among the plots in the current scan and those in the previous $N_{\rm scan}-1$ scans. To simplify exposition, assume that the current scan is $\ell=N_{\rm scan}$, so that the plot-lists $\bm{A}_{1},\ldots,\bm{A}_{N_{\rm scan}}$ are processed. Then, the trajectory of a prospective target is specified by an $N_{\rm scan}$-dimensional vector, say $\bm \xi=[\xi_1 \; \cdots \; \xi_{N_{\rm scan}}]\transp$, wherein $\xi_i=k$ means the target is observed at the $i$-th scan and the corresponding plot is the $k$-th row of $\bm S_i$. Accordingly, $\bigl[S_i(k,2)\; S_i(k,3)\; S_i(k,4)\bigr]\transp$ is the target position at time $S_i(k,5)$, and $S_i(k,1)$ is the statistic resulting from~\eqref{eq:filter_output}. Instead, $\xi_i=0$ means that the target is not observed at the $i$-th scan. The decision on the target presence is then made as
\begin{equation}
 \max_{\bm \xi\in \Xi} \mathcal T_{\bm \xi} \test \eta_{\rm TBD},
\end{equation}
where: $\eta_{\rm TBD}$ a threshold set to have desired probability of false alarm $P_{\rm fa}$; $\Xi$ is the set containing the vectors indexing all trajectories ending in a plot at scan $N_{\rm scan}$ and satisfying the required physical constraints on the target motion; and 
\begin{equation}
 \mathcal T_{\bm \xi} = \sum_{\substack{i=1,\; \xi_i\neq 0}}^{N_{\rm scan}} S_i(\xi_i,1),
\end{equation}
is the metric used to evaluate the trajectory index by $\bm \xi$.
If a target is declared, the associated trajectory is indexed by $\argmax_{\bm \xi \in \Xi} \mathcal T_{\bm \xi}$. It is worth mentioning that this approach can readily be generalized to the case where multiple targets can be present in the inspected volume $\mathcal{V}_{s}$, as shown in~\cite{Grossi-2013b, Aprile-2016}.

\section{Numerical analysis}\label{sec:Numerical analysis}
Here we report an example of application for the architecture in Fig.~\ref{fig:sm}: the considered setup and the corresponding performance are analyzed in Secs.~\ref{subsec_setting} and~\ref{subsec_num_{s}es}, respectively.

\subsection{System setup}\label{subsec_setting}
\begin{table}[t]
 \caption{System parameters}\label{tab:parameters}
 \begin{center}
    \renewcommand{\arraystretch}{1.2} 
    \begin{tabular}{p{0.45\columnwidth}p{0.45\columnwidth}}
        \toprule
         \rowcolor{Gray} $f_{o}=3.5$~GHz      & $\mathcal{P}= 4.803\upmu$W      \\
                         $W_{o}=15$~kHz       & $W_{\rm sub}= 720$~kHz      \\
         \rowcolor{Gray} $N_{o}=3300$         & $N_{\rm sub} = 32$          \\ 
                         $T_{o}=4.7623\upmu$s & $D_{\rm ris}= 8 \times 8=64$ \\
         \rowcolor{Gray} $N_{\rm sym} = 64$   & $D_{\rm tx}=5 \times 3=15$  \\
                          $B_{\rm sym}=76$    & $D_{\rm tx}=5 \times 3=15$ \\ 
        \bottomrule    
    \end{tabular}
    \end{center}
\end{table}

We consider the system parameters in Table~\ref{tab:parameters} and assume half-wavelength inter-element spacing at all arrays. For the sensing function, this setting implies that the range and radial velocity resolution are $c_{o}/(2N_{\rm sub}W_{\rm sub}) \simeq 6.5$~m and $c_{o}/(2 N_{\rm sym}T_{\rm sym}f_{o}) \simeq 9.4$~m/s, respectively, the maximum range allowed by the cyclic prefix is $c_{o}T_{o}/2 \simeq 713.7$~m, and the non-ambiguous range and radial velocity are $c_{o}/(2W_{\rm sub}) \simeq 208.1$~m and $c_{o}/(4T_{\rm sym}f_{o}) \simeq299.7$~m/s, respectively, where $c_{o}$ is the speed of light.

The BS is located at $[-1.5\; 1.5\; 25]\transp$, with the Cartesian coordinates expressed in meters, and its transmit and receive arrays are parallel to the ($x,z$)-plane and oriented towards the negative $y$ axis. Also, we assume equal power allocation across subcarriers, whereby $\gamma_{1}=\cdots\gamma_{N_{\rm sub}}=\gamma$. Finally, channel-matched beamforming is employed, whereby $\bm f_{q,c}$ and $\bm f_{q,s}$ are proportional to $\bm{h}_{q,c}$ and the right eigenvector of $\bm{G}_{q,\rm tx}$, corresponding to the largest eigenvalue, respectively.

The user is randomly dropped inside the rectangular cuboid $[20~\mathrm{m},40~\mathrm{m}]\times[-40~\mathrm{m},-20~\mathrm{m}]\times[1.5~\mathrm{m},2~\mathrm{m}]$, with its antenna parallel to the $(x,z)$-plane and oriented towards the positive $y$ axis. A line-of-sight channel is assumed, whereby we have
\begin{equation}
\bm{h}_{q,c} = \sqrt{\text{G}_{\rm{tx}}(\bm{\theta}_{\rm{u}})\text{G}_{\rm{u}}(\bm{\varphi}_{\rm{u}})}\frac{c_{o}}{4\pi d_{\rm{u}}f_{q}}\e^{-\i2\pi d_{\rm{u}}f_{q}}\bm{t}_{q}(\bm{\theta}_{\rm{u}}),
\end{equation}
where $\text{G}_{\rm{tx}}(\bm{\theta}) = \text{G}_{\rm{u}}(\bm{\theta})=\pi \cos(\theta^{\rm{az}})\cos(\theta^{\rm{el}})$ is the antenna gain for the BS and the user, $\bm \theta_{\rm{u}}$ is the angle of the departure, $\bm \varphi_{\rm{u}}$ is the angle of arrival and $d_{\rm{u}}$ is the distance. Finally, the noise variance of the receiver is $\sigma^{2}_{c}=1.918\times 10^{-16}$~W.

The RIS is located at $[0\;0\;25]\transp$, with the Cartesian coordinates expressed in meters, parallel to the $(x,z)$-plane and oriented towards the positive $y$ axis. The RIS is in close proximity of the BS to reduce the multiplicative path loss along the indirect path, and the forward link between the $j$-th transmitting element and the $i$-th RIS element on the $q$-th subcarrier is modeled as~\cite{Buzzi-2022}
\begin{align}
\bigl[\bm{G}_{q,\rm tx}\bigr]_{i,j} =& \sqrt{\text{G}_{\rm{tx}}(\bm{\theta}_{i,j})\text{G}_{\rm{ris}}(\bm{\varphi}_{i,j})}\frac{c_{o}}{4\pi d_{i,j}f_{q}}\e^{-\i2\pi d_{i,j}f_{q}},
\end{align}
where $\text{G}_{\rm{ris}}(\bm{\theta})=\pi \cos(\theta^{\rm{az}})\cos(\theta^{\rm{el}})$ is the element gain for the RIS, $\bm{\theta}_{i,j}$ is the angle of departure, $\bm{\varphi}_{i,j}$ is the angle of arrival, and $d_{i,j}$ is the distance. $\bm{G}_{q,\rm rx}$ is similarly modeled. For the monitored volume $\mathcal{V}_{s}$, we have $[R_{\min},R_{\max}]=[10~\mathrm{m},200~\mathrm{m}]$, $[\theta_{\min}^{\rm{az}},\theta_{\max}^{\rm{az}}]=[-22.5^\circ, 22.5^\circ]$, and $[\theta_{\min}^{\rm{el}},\theta_{\max}^{\rm{el}}]=[5^\circ, 15^\circ]$. This volume is scanned via $N_{\rm dir}=6$ illuminations with pointing directions $\bar{\bm{\theta}}_{1}=[-18.75^\circ\;10^\circ]\transp$, $\bar{\bm{\theta}}_{2}=[-11.25^\circ\;10^\circ]\transp$, $\bar{\bm{\theta}}_{3}=[-3.75^\circ\;10^\circ]\transp$, $\bar{\bm{\theta}}_{4}=[3.75^\circ\;10^\circ]\transp$, $\bar{\bm{\theta}}_{5}=[11.25^\circ\;10^\circ]\transp$, and $\bar{\bm{\theta}}_{6}=[18.75^\circ\;10^\circ]\transp$.

The prospective target has a random trajectory with a constant velocity. Also, a Swerling~I fluctuation is assumed, whereby $\alpha$ is modeled as a complex circularly-symmetric Gaussian random variable with variance dictated by the radar equation and equal to $ \sigma_{\text{RCS}} \text{G}^2_{\rm ris} (\bm{\phi})c_{o}^2/ ((4\pi)^3d^4 f_{o}^2)$, where $\bm{\phi}$ and $d$ are the target's angle and distance from the RIS, and $\sigma_{\text{RCS}}$ is the target radar cross-section (RCS). The RCS is set to have a \emph{nominal} SNR in~\eqref{eq:SNR_{s}} equal to 27~dB at the current scan when all power is devoted to the sensing function. 

At the radar receiver, we assume a maximum target speed of $40$~m/s and a noise variance of $\sigma^{2}_{s}=1.918\times 10^{-16}$~W. 
In the detector and plot-extractor, there are $N_{\rm del} N_{\rm dop}$ correlators covering the inspected delay and Doppler region, with $N_{\rm del}=60$ and $N_{\rm dop}=9$, and $\eta_{\rm plot}$ is set to have an average number of $6$ plots per scan under $\mathcal H_0$. In the TBD processor, the number of scans $N_{\rm{scan}}$ varies in the set $\{1, 5, 8, 12, 15\}$, a polynomial regression is used to smooth the trajectory of a detected target, and $\eta_{\rm TBD}$ is set to have $P_{\rm fa} =10^{-3}$. 

\begin{figure}[!t]
    \centering
    \includegraphics[width=0.95\columnwidth]{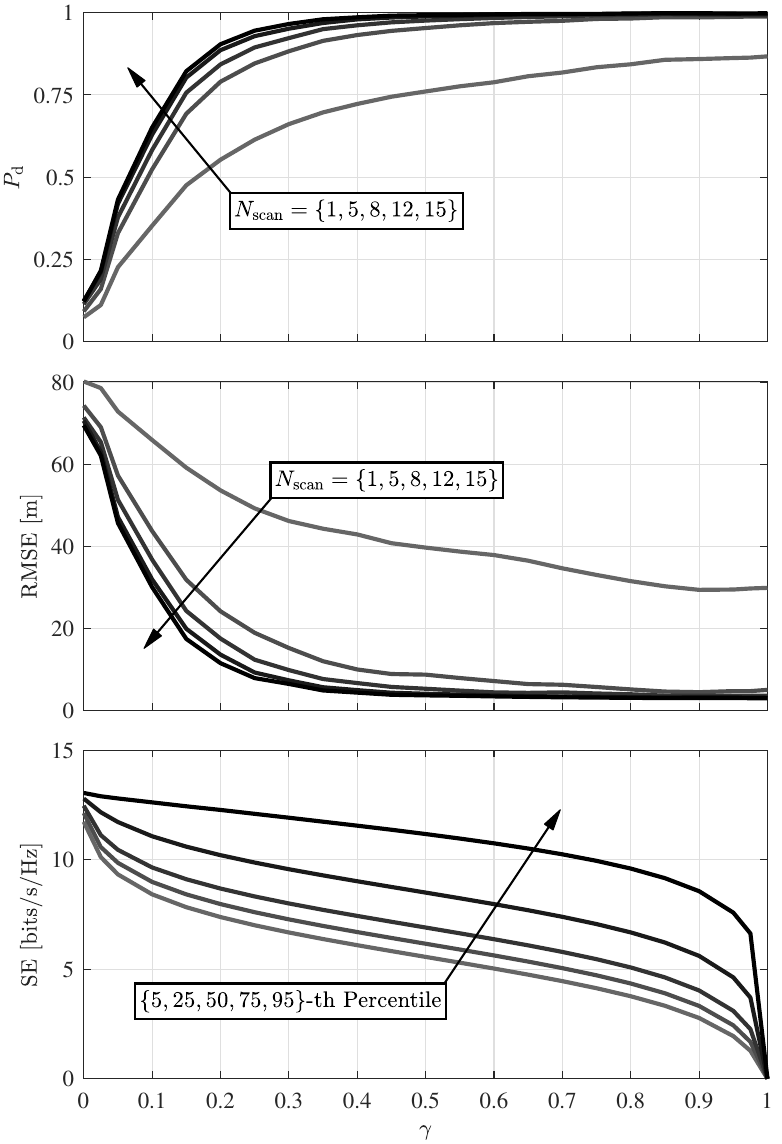}
    \caption{Sensing: $P_{\rm d}$ (top) and RMSE (middle) versus  $\gamma$  for various $N_{\rm{scan}}$. Communication: various percentiles of the user SE (bottom) versus  $\gamma$.}
    \label{fig:pd_{s}ate_gamma}
\end{figure}

\subsection{Numerical results}\label{subsec_num_{s}es}
The sensing performance is assessed in terms of the probability of detection\footnote{A detection is counted here if the hypothesis $\mathcal{H}_1$ is declared when $\mathcal{H}_1$ is true \emph{and} the position of the detected target lies within a sphere of radius 13 meters centered at the location of the true target.} ($P_{\rm d}$) and the root-mean-square error (RMSE) in the estimation of the target position in the current scan under $\mathcal{H}_1$. Instead, the communication performance is assessed in terms of the user SE in~\eqref{eq:se}; since the SE depends on the user location in the considered rectangular cuboid, we report its $\{5,25,50,75,95\}$-th percentiles. Fig.~\ref{fig:pd_{s}ate_gamma} reports all performance metrics versus the fraction of power dedicated to the sensing function. As expected, increasing $\gamma$ improves the target detection and estimation performance at the price of deteriorating the user SE. More interestingly, for a fixed $\gamma$, the target detection and estimation performances depend on the number of processed scans; for example, when $\gamma=0.2$, $P_{\rm d}$ increases from 0.53 to 0.90 as $N_{\rm scan}$ rises from 1 to 15. Notice that the performance gain granted by the multi-frame processing is more evident when $N_{\rm scan}$ is increased from $1$ to $5$ and substantially saturates beyond $N_{\rm scan}=12$; hence, a limited increase of the implementation complexity of the radar detector is sufficient to get a significant benefit. Overall, using a multi-frame detector allows for more favorable trade-offs between the sensing and communication functions.

\section{Conclusions}\label{sec:conclusions}
In this study, we have considered a BS that serves a downlink communication user and monitors a portion of the sky via a closely spaced passive RIS. Key design variables are the fraction of power allocated to the sensing and communication functions, the RIS response, and the implementation of the radar detector. Within this framework, we have designed the RIS response to scan the volume of interest. Additionally, we have proposed employing a multi-frame radar detector to exploit time correlation across scans; specifically, a TBD processor is employed to cope with the fast motion of an airborne target. Our main result is that using a multi-frame radar detector allows for more favorable sensing and communication trade-offs. Notably, increasing the number of scans processed by the radar detector allows for a reduction in the fraction of power dedicated to the sensing function without compromising the detection and estimation performance; the excess power can be redirected toward the user, thereby increasing its SE.

\end{document}